\documentclass[12pt]{article}
\usepackage{amssymb}
\usepackage{amsmath}
\usepackage{graphicx}
\usepackage{latexsym}

\begin{document}

\begin{center}
\vspace{1cm}{\Large {\bf Remarks on E11 approach}}

\vspace{1cm} {\bf H. Mkrtchyan} \footnote{ E-mail:hike@r.am} and
{\bf R. Mkrtchyan} \footnote{ E-mail: mrl@r.am} \vspace{1cm}

\vspace{1cm}

{\it Theoretical Physics Department,} {\it Yerevan Physics
Institute}

{\it Alikhanian Br. St.2, Yerevan 375036, Armenia}
\end{center}

\vspace{1cm}
\begin{abstract}

We consider a few topics in $E_{11}$ approach to
superstring/M-theory: even subgroups ($Z_2$ orbifolds) of $E_{n}$,
n=11,10,9 and their connection to Kac-Moody algebras; $EE_{11}$
subgroup of $E_{11}$ and coincidence of one of its weights with
the $l_1$ weight of $E_{11}$, known to contain brane charges;
possible form of supersymmetry relation in $E_{11}$; decomposition
of $l_1$ w.r.t. the $SO(10,10)$ and its square root at first few
levels; particle orbit of $l_1 \ltimes E_{11}$. Possible relevance
of coadjoint orbits method is noticed, based on a self-duality
form of equations of motion in $E_{11}$.

\end{abstract}

\renewcommand{\thefootnote}{\arabic{footnote}}
\setcounter{footnote}0 {\smallskip \pagebreak }

\section{Introduction}

One of the recent ideas on a hidden structures in M-theory is that
of a hidden $E_{11}$ and/or $E_{10}$ Kac-Moody Lie algebra
symmetry \cite{w0,w00,n4}, generalizing an U-duality of
compactified superstrings/M-theory. Appearance of E series of Lie
algebras as a symmetry of supergravity theories started from the
discovery of $E_7$ as a symmetry (of equations of motion) of
maximal four-dimensional supergravity \cite{cremmer}. Afterwards,
$E_n$ type algebras, with Dynkin diagrams given below were
discovered to appear in compactifications of 11d supergravity and
superstrings to $11-n$ dimensions, including $E_9$ affine algebra
in 2d \cite{julia} and $E_{10}$ in a 1-dimensional reduction in a
form of a particle motion in an $E_{10}$ Weyl chamber (see
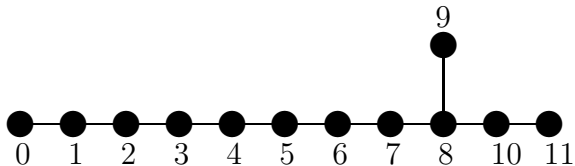
\cite{n3}).

\begin{align}\label{e11}
\begin{picture}(0,120)(0,-50)
\put(-120,0){\line(1,0){200}}\put(-120,0){\circle*{10}}
\put(-100,0){\circle*{10}} \put(-80,0){\circle*{10}}
\put(-60,0){\circle*{10}} \put(-40,0){\circle*{10}}
\put(-20,0){\circle*{10}} \put(0,0){\circle*{10}}
\put(20,0){\circle*{10}} \put(40,0){\circle*{10}}
\put(40,0){\line(0,1){30}} \put(40,30){\circle*{10}}
\put(60,0){\circle*{10}}\put(80,0){\circle*{10}}
\put(-122,-15){\text{0}}
\put(-102,-15){\text{1}}\put(-82,-15){\text{2}}
\put(-62,-15){\text{3}}\put(-42,-15){\text{4}}
\put(-22,-15){\text{5}}\put(-2,-15){\text{6}}
\put(18,-15){\text{7}}\put(38,-15){\text{8}}
\put(58,-15){\text{10}}\put(78,-15){\text{11}}
\put(37,37){\text{9}} \put(-180, -50){\text{Dynkin diagrams of
$E_{12-n}$ are given by nodes n, n+1, ..., 11}}
\end{picture}
\end{align}

It was the main idea of \cite{w0}  to consider $E_{11}$ as a
symmetry of M-theory. Formally, its Dynkin diagram appears in
U-duality considerations of M-theory compactified to 0 dimensions
\cite{pioline}. The point is that $E_{11}$ was suggested in
\cite{w0} as a symmetry of opposite extreme - completely
uncompactified theory. There is a number of arguments in favor of
this idea: the field content of model of \cite{w0} recovers (the
first levels of) the M-theory, the T-duality between IIA and IIB
theories appears to be a simple property of $E_{11}$ Dynkin
diagram, the brane charges seem to fill in one of the fundamental
representations of $E_{11}$, and others. The accompanied
difficulties can be seen from this last observation: since in the
usual approach the space-time is associated with point-like
charge, which now is the part of $E_{11}$ irrep, the $E_{11}$
covariance requires substitution of space-time with duals of all
(infinite number of) brane charges, which apparently is not a
standard situation in field or string theories.

The similar $E_{10}$ suggestion \cite{n4} is much more compact and
more precise. It deals with 1d sigma model, instead of
infinite-dimensional one, namely that based on a coset of $E_{10}$
group, with fields depending on one parameter in a
reparametrization invariant way. The price is the loss of (at
least explicit) Lorentz invariance, since the space-time is
introduced in this approach by assumption that coefficients of
expansion of fields over coordinates appear on the higher levels
of algebra (the number of which is infinite, due to the Kac-Moody
nature of $E_{10}$).

The $E_{11}$ model will be our main object of study in this paper,
we will present a few results on different aspects of the topic.

According to \cite{w0,w00}, the hypothesis is that sigma model
over coset space $E_{11}/K_{11}$ gives some description of
M-theory. To define $K_{11}$ we first introduce notations for
Kac-Moody Lie algebra with Cartan matrix $A_{ij}$. It's the Lie
algebra with generators $h_i, e_i, f_i$ and relations:

\begin{eqnarray} \label{km}
[h_i,e_j]&=& A_{ij} e_j  \\\label{2} [h_i,f_j]&=&-A_{ij}f_j
\\\label{3}
[e_i,f_j]&=& \delta_{ij}h_j \\ \label{4}
ad(e_i)^{(1-A_{ij})}e_j&=&0, \\
\label{5}ad(f_i)^{(1-A_{ij})}f_j&=&0
\end{eqnarray}

All other generators should be obtained from these prime
generators by all possible multiple commutators, factorized over
relations (\ref{km})-(\ref{5}). Matrix $A_{ij}$ has the following
properties: $A_{ii}=2$, $A_{ij}$ are non-positive integers such
that from $A_{ij}=0$ follows $A_{ji}=0$. $A_{ij}$ alternatively
and equivalently can be presented by Dynkin diagram, with simple
rules of equivalence. For example, for symmetric $A_{ij}$ with
non-diagonal entries 0 or -1 Dynkin diagram is given by nodes
equal to dimensionality of $A_{ij}$, nodes $i$ and $j$ are
connected by simple line if and only if $A_{ij}=-1$.

Next, Chevalley involution of an arbitrary Kac-Moody algebra is
the Lie algebra automorphism

\begin{eqnarray} \label{chev}
h_i\rightarrow -h_i \\
e_i\rightarrow -  f_i \\
f_i\rightarrow -e_i
\end{eqnarray}

Subalgebra $K_{11}$  consists of generators invariant under
Chevalley involution, it is generated by elements $e_i-f_i$. $K_n$
is a key object for the discussion of supersymmetry, also, see
below. The study of $K_9$ see in \cite{n6}.

In the body of paper, in Section 2 we consider the even ($Z_2$
invariant, see (\ref{z2})-(\ref{z2z})) subalgebras of $E_n$
algebras. From the M-theory viewpoint they are relevant,
particularly, for orbifold considerations \cite{ganor}. Moreover,
for finite dimensional algebras $E_7, E_8$ the even subalgebras
actually coincide with Chevalley-invariant subalgebras $K_7, K_8$.
This statement actually extends to all finite-dimensional algebras
where corresponding $K_n$ has rank $n$. In \cite{mrl} it was
suggested that this coincidence extends to Kac-Moody algebras
$E_n, n=9,10,11$, however, it seems that  this assumption is not
correct. Which concerns a description of even subalgebras, our
claim is that the even roots of $E_{11}, E_{10}$ and $E_9$
coincide with all roots of $EE_{11}$, $DE_{10}$ and $D_8^{(1)}$,
respectively. For $E_{11}$ that is supported by numerical
calculations, for $E_{10}$ and $E_{9}$ that is proved below, for
$E_{9}$ that gives the complete coincidence of algebras, since the
multiplicities of imaginary roots coincide, also.

Next Section 3 considers the weights of fundamental
representations of $EE_{11}$, introduced in previous Section 2.
Observation is that one of its fundamental weights coincides with
weight of $l_1$ - first fundamental weight of $E_{11}$, known to
contain the brane charges \cite{w3}. This kind of considerations
are aimed to discussion of possible supersymmetry relation in
$E_{11}$ theory:

\begin{eqnarray}\label{susy}
\left\{Q,Q\right\}=Z
\end{eqnarray}

In usual supersymmetric theories the supercharges $Q$ should be a
representations of both compact subgroup $K_n$ and Lorentz group.
E.g., 3d compactification of 11d supergravity gives a $E_8/SO(16)$
supersymmetric sigma-model \cite{sch}, with supercharges in spinor
representation of Lorentz group and vector of SO(16). In $E_{11}$
approach these two groups are joined into $K_{11}$ group. From the
other side, the anticommutator of supercharges gives the brane
charges $Z$, which, as argued in \cite{w3}, is $l_1$, the
irreducible representation of $E_{11}$. So, roughly speaking, the
symmetric square of $Q$ representation of $K_n$ gives a
fundamental irrep of $E_{11}$. More precisely, one can imagine
that some Klebsh-Gordon coefficients can enter in r.h.s. of
(\ref{susy}), so $l_1$ is one of irreps, appearing in the r.h.s
after decomposition to irreducible representations. The symmetric
square of highest-weight representation gives, particularly, the
irrep with doubled highest weight. So, we see that the coefficient
2 is missing in abovementioned statement of coincidence of one of
weights of $EE_{11}$ with weight of $l_1$, saying least.
Nevertheless, the coincidence of highest weights can signal on
some relevance of $EE_{11}$. The attempt to introduce a
supersymmetry in the $E_{11}$ approach was done in \cite{igor},
where the Killing spinor equations are constructed, and fermionic
generators are introduced into a part of $G_{11}$ algebra, which
was an intermediate step in construction of $E_{11}$ in \cite{w0}.
It seems, however, that these results are not relevant for
supersymmetrization of $E_{11}$ itself, (\ref{susy}), due to the
few reasons, one of which is that group, considered in
\cite{igor}, includes momenta which is not the part of $E_{11}$.
Further discussion of problems of \cite{igor} see in Section 4

The discussion of supposed susy relation (\ref{susy}) is continued
in the next Section 4, where we study the expansion of first
fundamental weight $l_1$ of $E_{11}$ w.r.t. the levels of root
$e_{11}$ of $E_{11}$, (rightmost root of diagram (\ref{e11})) and
show that the subset of representations at first three levels can
be obtained as a symmetric square of representations of
corresponding compact subgroup $SO(10) \times SO(10)$. This result
is another face of similar phenomena found in \cite{w6}.

Section 5 is devoted to the study of hypothesis that finally
symmetry group  should be extended to semidirect product of $l_1$
and $E_{11}$ \cite{w2}. We calculate the little group for particle
orbit, i.e. for a given point in the space $l_1$ of brane charges,
when all charges are zero, except the particle one, we calculate
its stabilizer in $E_{11}$. It appears to have an explicit
description in terms of basic generators of $E_{11}$.

Conclusion contains the resume of results and ways of their
development, particularly, possible relevance of coadjoint orbits
of $E_{11}$ is discussed.

\section{Even subalgebras of $E_n$}

We consider involution of $E_n$ algebras, given by

\begin{eqnarray} \label{z2}
h_i\rightarrow h_i \\
e_i\rightarrow -  e_i \\
f_i\rightarrow -f_i \label{z2z}
\end{eqnarray}

The corresponding invariant subalgebra is given by generators
$h_i$ and those of even power of $e_i$ or $f_i$. Let's denote that
by $Z_2(E_n)$. Study of this subalgebras is relevant for $Z_2$
orbifolds \cite{ganor}. For finite dimensional Lie algebras $g$
with the property that rank of CSA of $K_n$ (Chevalley-invariant
subalgebra) is maximal, i.e. equal to n, $K_n$ coincide with
$Z_2(g)$,

\begin{eqnarray} \label{kz}
K_n(g) = Z_2(g)
\end{eqnarray}

It is suggested in \cite{mrl} that (\ref{kz}) extends to Kac-Moody
algebras. However, the problem is in Cartan subalgebra. Although
one can find \cite{mrl} a lot of commuting generators, even with
hermiticity properties, their diagonalizability is questionable,
since one can show that they have to be diagonalized in the
infinite subspaces \footnote{We are indebted to H.Nicolai for
e-mail correspondence stressing the importance of
diagonalizability of Cartan generators}.

As proposed in \cite{w6} and  \cite{kn}, the study of bilinear
invarianrt forms can shed light on a problem of connection of
compact subalgebra with Kac-Moody algebras. Particularly, in
\cite{kn} is shown, that special contravariant Hermitian bilinear
form is positively defined on $K_n$, and this leads to a
conclusion that $K_n$ is not a semisimple Kac-Moody algebra.

Now we shall try to construct even roots of $E_{11}$ from some
basic even roots. We would like to introduce the subalgebra of
$E_{11}$ generated by CSA and following generators (Lie algebra
commutators are implied) and their opposite roots partners:

\begin{eqnarray} \label{ee11g}
a_1= e_7 e_8 e_9 e_{10}, a_2=e_1 e_2,  a_3=e_3 e_4, a_4=e_5 e_6,\\
\nonumber a_5=e_7 e_8, a_6=e_{10} e_{11}, a_7=e_8 e_9, a_8=e_6
e_7,\\ \nonumber a_9=e_8 e_{10},a_{10}=e_4 e_5,  a_{11}=e_2 e_3,\\
\nonumber
\end{eqnarray}
Definition of $a_1$ is actually unique, up to overall sign, since
although Lie brackets can be arranged in different ways, results
coincide. Roots of (\ref{ee11g}) are real. One can find the
corresponding Cartan matrix and Dynkin diagram:

\begin{eqnarray}\label{o11a}
EE_{11}=
\begin{array}{*{20}c}
 2&0&0&-1&0&0&0&0&0&0&0 \\
    0&2&-1&0&0&0&0&0&0&0&0 \\
    0&-1&2&-1&0&0&0&0&0&0&0 \\
    -1&0&-1&2&-1&0&0&0&0&0&0 \\
    0&0&0&-1&2&-1&0&0&0&0&0 \\
    0&0&0&0&-1&2&-1&0&0&0&0 \\
    0&0&0&0&0&-1&2&-1&0&0&0 \\
    0&0&0&0&0&0&-1&2&-1&-1&0 \\
    0&0&0&0&0&0&0&-1&2&0&0 \\
    0&0&0&0&0&0&0&-1&0&2&-1 \\
    0&0&0&0&0&0&0&0&0&-1&2
\end{array}
\end{eqnarray}

\begin{align}\label{ee11}
\begin{picture}(0,100)(0,-50)
\put(-80,0){\line(1,0){160}}\put(-80,0){\circle*{10}}
\put(-60,0){\circle*{10}} \put(-40,0){\circle*{10}}
\put(-40,0){\line(0,1){30}} \put(-40,30){\circle*{10}}
\put(-20,0){\circle*{10}} \put(0,0){\circle*{10}}
\put(20,0){\circle*{10}} \put(40,0){\circle*{10}}
\put(40,0){\line(0,1){30}} \put(40,30){\circle*{10}}
\put(60,0){\circle*{10}}\put(80,0){\circle*{10}}
\put(60,0){\circle*{10}}\put(80,0){\circle*{10}}
\put(-82,-15){\text{2}}
\put(-62,-15){\text{3}}\put(-42,-15){\text{4}}
\put(-22,-15){\text{5}}\put(-2,-15){\text{6}}
\put(18,-15){\text{7}}\put(38,-15){\text{8}}
\put(56,-15){\text{10}}\put(78,-15){\text{11}}\put(37,37){\text{9}}
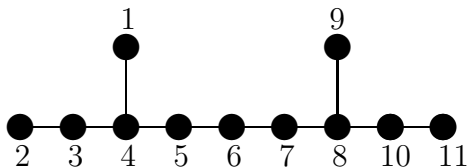
\put(-42,37){\text{1}} \put(-80, -50){\text{Dynkin diagram of
$EE_{11}$ algebra}}
\end{picture}
\end{align}

where simple roots in (\ref{ee11}) are enumerated in agreement
with (\ref{o11a}). One can construct the corresponding abstract
Kac-Moody algebra, we denote it by $EE_{11}$ since it contains two
E type tails, and this notation is similar to that of hyperbolic
algebras - AE, BE, CE, DE. Since roots (\ref{ee11g}) are real,
according to \cite{nic2}, algebra (\ref{ee11}) is isomorphic to
above defined subalgebra in $E_{11}$. Our hypothesis is that this
algebra has the same roots as $Z_2(E_{11})$, i.e. all even roots
of $E_{11}$. Statement seems to be simple, nevertheless, we were
not able to prove it algebraically, due to unknown structure of
roots system. Instead we checked that up to level 146 by the help
of computer program (available upon request), which generates the
roots for an arbitrary input Dynkin diagram. The number of roots
up to the level 146 (inclusively) is 19661788 (without counting
multiplicity), so coincidence is considerable. Since multiplicity
of real roots is one, this statement means that these two algebras
coincide at least in the sector of real roots. However, as shown
in \cite{ganor} for similar considerations for $E_{10}$ (see
below), for imaginary roots there is difference in multiplicities
in the case of $Z_2(E_{10})$ and $DE_{10}$, so we are confident in
the same statement for $E_{11}$.

Similar statements (on the coincidence of even roots and roots of
subalgebras of simple even roots) for $E_{10}$ and $E_{9}$ can be
proved. Corresponding composite roots (generators) and Dynkin
diagram for $Z_2(E_{10})$ are:

\begin{eqnarray} \label{de10-1}
a_1=e_6 e_7 e_8 e_9, a_2=e_2 e_3, a_3=e_4 e_5, a_4=e_6 e_7,
a_5=e_9 e_{10}, \\a_6=e_7 e_8, a_7=e_5 e_6, a_8=e_7 e_9, a_9=e_3
e_4, a_{10}=e_1 e_2\nonumber
\end{eqnarray}

\begin{align}\label{de10}
\begin{picture}(0,100)(0,-50)
\put(-80,0){\line(1,0){140}}\put(-80,0){\circle*{10}}
\put(-60,0){\circle*{10}} \put(-40,0){\circle*{10}}
\put(-60,0){\line(0,1){30}} \put(-60,30){\circle*{10}}
\put(-20,0){\circle*{10}} \put(0,0){\circle*{10}}
\put(20,0){\circle*{10}} \put(40,0){\circle*{10}}
\put(20,0){\line(0,1){30}} \put(20,30){\circle*{10}}
\put(60,0){\circle*{10}} \put(-82,-15){\text{2}}
\put(-62,-15){\text{3}}\put(-42,-15){\text{4}}
\put(-22,-15){\text{5}}\put(-2,-15){\text{6}}
\put(18,-15){\text{7}}\put(38,-15){\text{9}}
\put(58,-15){\text{10}}\put(17,37){\text{8}}
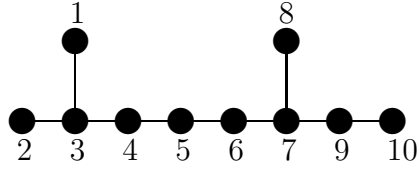
\put(-62,37){\text{1}} \put(-80, -50){\text{Dynkin diagram of
$DE_{10}$ algebra}}
\end{picture}
\end{align}
\vspace{5mm}

For $E_{9}$:

\begin{eqnarray} \label{d91c}
a_1=e_1 e_2, a_2=e_5 e_6 e_7 e_8, a_3=e_3 e_4, a_4=e_5 e_6,\\
a_5=e_8 e_9, a_6=e_6 e_7, a_7=e_4 e_5, a_8=e_6 e_8, a_9=e_2 e_3
\nonumber
\end{eqnarray}

\begin{align} \label{d81}
\begin{picture}(0,100)(0,-50)
\put(-60,0){\line(1,0){120}} \put(-60,0){\circle*{10}}
\put(-40,0){\circle*{10}} \put(-40,0){\line(0,1){30}}
\put(-40,30){\circle*{10}} \put(-20,0){\circle*{10}}
\put(0,0){\circle*{10}} \put(20,0){\circle*{10}}
\put(40,0){\circle*{10}} \put(40,0){\line(0,1){30}}
\put(40,30){\circle*{10}} \put(60,0){\circle*{10}}
\put(-62,-15){\text{2}}\put(-42,-15){\text{3}}
\put(-22,-15){\text{4}}\put(-2,-15){\text{5}}
\put(18,-15){\text{6}}\put(38,-15){\text{7}}
\put(58,-15){\text{9}}\put(37,37){\text{8}} \put(-42,37){\text{1}}
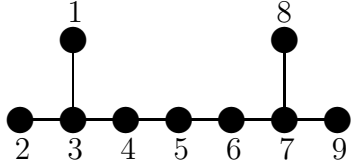
\put(-80, -50){\text{Dynkin diagram of $D_8^{(1)}$ algebra}}
\end{picture}
\end{align}

Both for $E_{10}$ and $E_9$ coincidence of roots follows from two
facts. First, root lattices coincide \cite{ganor}, namely, even
root sublattice of $E_{10}$ ($E_9$) coincide with lattice of
$DE_{10}$ (\ref{de10})   ($D_8^{(1)}$ (\ref{d81})). Second,
description of all roots is given in (\cite{kac}, p.67), in terms
of lattices - all real roots are all those elements of lattice
with square equal two, and all other roots (i.e. imaginary ones)
are all those elements of lattice with square less or equal to
zero. For $E_9$ that means complete coincidence of algebras:

\begin{eqnarray} \label{kz2}
Z_2(E_9)=D_8^{(1)}
\end{eqnarray}
since multiplicities of imaginary roots coincide (multiplicities
of $Z_2(E_9)$ are that of $E_9$, which is 8, since that is an
affine algebra $E_8^{(1)}$, and multiplicity of roots of
$D_8^{(1)}$ is 8, also).

For $E_{10}$ the problem of multiplicities is more complicated,
and in \cite{ganor} it is shown that actually multiplicities of
$Z_2(E_{10})$ and $DE_{10}$ are different. It would be interesting
to describe that difference explicitly.

For $E_{11}$ the coincidence of its even lattice of and that of
$EE_{11}$ can be proved, also, but it is not enough for
coincidence of all roots, since for these non-hyperbolic algebras
there is no similar description of roots.

Coincidence of even roots of $E_{10}$ and all roots of $DE_{10}$
also follows from statements on a level decompositions of these
two algebras, proved in Section 4.2 of \cite {kn2}.

\section{On a representations of $EE_{11}$}

The weights of the fundamental representations of $E_{11}$. can be
obtained by the rows of an inverse Cartan matrix
$(EE_{11})^{(-1)}$ of (\ref{ee11}). We would like to compare these
weights with those of $E_{11}$:

\begin{eqnarray}\label{einv}
E_{11}^{-1}=\frac{1}{2} \left[ \begin{array}{*{20}c}
-1&0&1&2&3&4&5&6&3&4&2 \\
    0&0&2&4&6&8&10&12&6&8&4 \\
    1&2&3&6&9&12&15&18&9&12&6 \\
    2&4&6&8&12&16&20&24&12&16&8 \\
    3&6&9&12&15&20&25&30&15&20&10 \\
    4&8&12&16&20&24&30&36&18&24&12 \\
    5&10&15&20&25&30&35&42&21&28&14 \\
    6&12&18&24&30&36&42&48&24&32&16 \\
    3&6&9&12&15&18&21&24&11&16&8 \\
    4&8&12&16&20&24&28&32&16&20&10 \\
    2&4&6&8&10&12&14&16&8&10&4
\end{array} \right]
\end{eqnarray}

The only subtlety is that rows of inverse Cartan matrix express
weights in the basis of simple roots of a given algebra, so for
comparison we should express both in the same basis, e.g. in a
basis of simple roots of $E_{11}$. The expression of simple roots
of $EE_{11}$ through the simple roots of $E_{11}$ is given in
(\ref{ee11g}). So we should multiply the matrix $(EE_{11})^{(-1)}$
from the right by the transformation matrix

\begin{eqnarray}\label{t}
T=\left[ \begin{array}{*{20}c}
0&0&0&0&0&0&1&1&1&1&0 \\
    1&1&0&0&0&0&0&0&0&0&0 \\
    0&0&1&1&0&0&0&0&0&0&0 \\
    0&0&0&0&1&1&0&0&0&0&0 \\
    0&0&0&0&0&0&1&1&0&0&0 \\
    0&0&0&0&0&0&0&0&0&1&1 \\
    0&0&0&0&0&0&0&1&1&0&0 \\
    0&0&0&0&0&1&1&0&0&0&0 \\
    0&0&0&0&0&0&0&1&0&1&0 \\
    0&0&0&1&1&0&0&0&0&0&0 \\
    0&1&1&0&0&0&0&0&0&0&0
\end{array} \right]
\end{eqnarray}

and obtain

\begin{eqnarray}\label{oinvt}
(EE_{11})^{-1}T=\frac{1}{8} \left[ \begin{array}{*{20}c}
2&8&10&16&18&24&26&32&14&20&12 \\
    -4&0&4&8&12&16&20&24&12&16&8 \\
    0&8&8&16&24&32&40&48&24&32&16 \\
    4&16&20&32&36&48&60&72&36&48&24 \\
6&16&22&32&38&48&54&64&34&44&20\\
      8&16&24&32&40&48&56&64&32&40&16\\
      10&16&26&32&42&48&58&64&30&44&20\\
     12&16&28&32&44&48&60&72&36&48&24\\
      6&8&14&16&22&24&30&32&18&20&12\\
      8&8&16&16&24&32&40&48&24&32&16\\
      4&0&4&8&12&16&20&24&12&16&8\\
\end{array} \right]
\end{eqnarray}

The second row of (\ref{oinvt}) and first row of (\ref{einv})
coincide. As mentioned in the introduction, this statement has
some resemblance with the one that is implied by (supposed)
supersymmetry relation (\ref{susy}), although it differs in few
important points - coefficient 2 is missing, and, more important,
$EE_{11}$ is not a $K_{11}$, although can have similar properties.
It is also worth recalling the discussion of \cite{wit}, where the
traces of needed phenomena were noticed, namely at certain
dimensions certain brane charges are not only in the
representation of corresponding susy algebra's automorphism group,
but also combine into representations of U-duality group. For
example, consider maximal susy algebra in 4 dimensions, obtained
from reduction to 4d of 11d susy algebra. The corresponding 4d
algebra has SU(8) as an automorphism group. Scalar central charges
appear from a few sources: 7 from vector, 21 from membrane charge,
and 28 from five-brane charge, altogether they combine into 56 of
4d U-duality group $E_7$, which contains SU(8) as its maximal
compact subgroup. This field worths further study, the key element
should be the theory of $K_{11}$ group's representations.

\section{$l_1$ expansion over $SO(10,10)$}

Lets's consider the expansion of both sides of supposed susy
relation (\ref{susy}) over $SO(20)$ ($SO(10,10)$ if reality
properties taken into account) subgroup of $E_{11}$, obtained by
removing most right root (number 11 on diagram (\ref{e11}). The
corresponding compact group is $SO(10)\times SO(10)$. This
expansion can shed some light on whether such a relation can
exist, at least on a first few terms of expansion. The level
expansion of $l_1$ with respect to the generator $e_9$ was
suggested in \cite{w6},  where it was shown, that usual brane
charges($P_\mu, Z_{\mu\nu}, Z_{\mu_1\mu_2\mu_3\mu_4\mu_5}$) appear
on first three levels of that expansion. As is well-known, one can
fulfill relation (\ref{susy}) with Q as a spinors of corresponding
compact group $SO(11)$. We would like to consider the same problem
with expansion over $e_{11}$, with compact group $SO(10)\times
SO(10)$, which was not supported by existence of any
supersymmetric theories, but, from the other side, should exist,
provided $E_{11}$ hypothesis and relation (\ref{susy}) are
correct.

Expansion over $e_{11}$ goes with non-negative powers. This is
clear when recalling an approach of \cite{w6} - in that paper
$l_1$ is identified with the subspace of $E_{12}$, linear over
$e_0$. This subspace evidently is a representation of $E_{11}$,
and actually the first fundamental representation, with $e_0$ as a
highest weight vector. The same is true for the zeroth order over
$e_{11}$ -  it is a representation with highest vector $e_0$,
since all $f_i, i>0$ commute with $e_0$ and action of $h_i, i>0$
on $e_0$ gives the only non-zero Dynkin label $p_1=1$, i.e. that
of vector representation of SO(20), $\underline{20}$. Next is
linear over $e_{11}$ representation of SO(20). It is easy to
understand, that the highest vector for that representation is
unique, namely that given by unique nonzero commutator of
$e_0,e_1, ..., e_{11}$ (except $e_9$). The only nonzero Dynkin
label is $p_9=1$, i.e. that is one of two Weyl spinor
representations of SO(20), of dimensionality 512. Next is the
second order over $e_{11}$ representation, which is the last one
we are interested in. It can be shown to be fifth-rank
antisymmetric tensor $Z_5$, with highest vector $\sum n_i e_i$,
$n_0=...=n_5=1, n_6=2,n_7=3,n_8=4,n_9=2,n_{10}=3,n_{11}=2$. The
corresponding Dynkin label is $p_5=1$ (other $p_i=0$). We are
interested in decomposition of these few first level
representations over compact subgroup of SO(10,10). The
$\underline{20}$ is of course $\underline{10} + \underline{10}$,
$\underline{512}$ is
$(\underline{16},\underline{16})+(\underline{16},\overline{\underline{16}})$
(chirality of $SO(20)$ representation choose chirality of
$SO(10)$). $Z_5$ is decomposed in more or less clear way,
according to all possible choices of belonging of indexes to two
product SO(10) groups:
$Z_5=Z_{5,0}+Z_{4,1}+Z_{3,2}+Z_{2,3}+Z_{1,4}+Z_{0,5}=(\underline{252},\underline{1})+
(\underline{210},\underline{10})+(\underline{120},\underline{45})+
(\underline{45},\underline{120})+
(\underline{10},\underline{210})+(\underline{1},\underline{252})$
Moreover, fifth index tensors should be decomposed into their
chiral parts, i.e. into self-dual and anti-self-dual tensors of
dimensionality 136. Here we imagine complex field of coefficients,
neglecting reality properties of groups involved, their will tune
themselves according to the choice of Chevalley subgroup.

After this preparatory work we can look for an answer on a
possibility of taking a square root of $l_1$, i.e. finding a
representation of $K_{11}$, symmetric square of which contains
$l_1$. It appears that one can take the following combination of
representation of $SO(10)\times SO(10)$. This combination should
be considered as a decomposition of irrep of K we are seeking for,
w.r.t. the $SO(10)\times SO(10)$:
\begin{eqnarray} \label{q}
(\underline{16},\underline{1})+ (\underline{1},\underline{16})
\end{eqnarray}

Symmetric square of this representation is
\begin{eqnarray} \label{q1}
(\underline{10},\underline{1})+
(\underline{1},\underline{10})+(\underline{16},\underline{16})+
(\underline{136},\underline{1})+ (\underline{1},\underline{136}))
\end{eqnarray}
This includes representations of first level, second level, and
part of representations of third level, obtained above in
decomposition of $l_1$ w.r.t. the powers of $e_{11}$.  I.e. we
find the square root of (part of the) first few levels of $l_1$.
Finally, the susy relation, in this approximation, can be
represented in a well-known form

\begin{eqnarray}\label{susy2}
\left\{Q_\alpha,Q_\beta\right\}=Z_{\alpha\beta}
\end{eqnarray}

This is a standard form of 11d susy relation, where supersymmetry
charge is 32 dimensional, giving in the r.h.s all possible 528
central charges. That means that we are dealing with the same
$SL(32)$ invariant susy algebra, decomposed with respect to
different subalgebras. The natural question is whether there exist
$SO(10) \times SO(10)$ covariant supergravity theories, with susy
algebra (\ref{susy2}). The $SL(32)$ invariance of (\ref{susy2})
doesn't persist on the higher levels, since $E_{11}$ does not have
such a subgroup (see discussion in \cite{k1}, where it was shown
that antisymmetric tensor representations, precisely corresponding
to those of $SL(32)$ decomposed w.r.t. the $SL(11)$ can be
identified on a first 4 levels of $K_{11}$).

Let's mention a difference of previous considerations with
approach of \cite{igor}. First, as mentioned in the Introduction,
the groups considered are different - $G_{11}$ of \cite{igor}
includes momenta $P\mu$, which is not the part of $E_{11}$.
Moreover, it is argued in \cite{igor} that two spinor generators
are needed in supersymmetrization of $G_{11}$, on the basis that
11d conformal group $SO(2,11)$ should be, finally, part of the
symmetries considered. That statement stress that $SO(2,11)$ group
is not part of $E_{11}$, from the same fact of absence of momenta
$P_\mu$ in $E_{11}$ (as well as its conformal counterpart
$K_\mu$).

This discussion stress the problem of finding the generalization
of $E_{11}$, which will include momenta and possibly the whole
conformal group $SO(2,11)$. As mentioned in Introduction, the
generalization including momenta,  actually the whole multiplet
$l_1$, is suggested in \cite{w6} as semidirect product of $l_1$ on
$E_{11}$. Inclusion of conformal SO(2,11) will require further
extension of this group.

Next remark concerns the possibility of continuation of above
analysis to higher levels of $e_{11}$. The problem is in that
there is no corresponding grading of desired representation of
$K_{11}$. So it is not clear how to continue finding next terms of
decomposition of that representation w.r.t. the $SO(10) \times
SO(10)$.

It is worth mentioning here the connection between $SO(10) \times
SO(10)$ subgroups of $SO(20)\subset E_{11}$, with $EE_{11}$. From
the diagram of $EE_{11}$ one can easily read off two SO(10)
subgroups, constructed from two $D_5$ sub-diagrams, that appear
after removing the middle root (number six). One can easily
understand, that they are the same $SO(10)$ subgroups of $SO(20)$
and $E_{11}$. It follows from the fact that eleventh root e11 of
$E_{11}$ enters in the sixth root of $EE_{11}$, only, and from the
fact mentioned in Section 1, that compact subgroup of $D_{10}$ can
be represented as its even subgroup.

\section{Particle orbit in $l_1 \ltimes E_{11}$}

In the search of a space-time in the $E_{11}$ approach, West
\cite{w6} suggested to extend the symmetry group to semidirect
product $l_1\ltimes E_{11}$, which is similar to Poincare group.
Then one has to consider a unitary representation of this group,
which can be constructed by Wigner's little group method, which is
recently applied to construction of irreps of semidirect product
of Lorentz and tensorial translations group \cite{mrl2}.  The
method requires choosing of the orbit of action of $E_{11}$ on
$l_1$, and then construction of unitary irreps of little group -
stabilizer of a given point of the orbit in the $E_{11}$. We will
apply this method to particle orbit.

According to suggestion \cite{w6} representation $l_1$ contains
all brane charges. Particularly,  the decomposition of $l_1$
w.r.t. the $SL(11)$ subgroup of $E_{11}$ starts from vector
representation $P_\mu$. Particle can be naturally defined as a
configuration of brane charges when all of them are zero, except
$P_\mu$. Note that we are dealing with general linear $GL_{11}$
group, which gives usual Lorentz SO(11) after taking a
Chevalley-invariant (=compact) subgroup of $GL_{11}$.
Correspondingly, an arbitrary vector $P_\mu$ can be transformed
into any other vector, so we can choose

\begin{eqnarray}\label{pmu}
P_\mu=(1,0,0,...)
\end{eqnarray}

Next, our aim is to define an orbit of this point under action of
the whole group $E_{11}$. Desired orbit is a factor of $E_{11}$
over $L$, the stabilizer of $P_\mu$, i.e. subgroup of $E_{11}$,
which leaves $P_\mu$ unchanged. So, our task is to find the
subgroup L. Note that $P_\mu$ (\ref{pmu}) is represented by just
$e_0$. So, we are seeking a stabilizer of $e_0$ in $E_{11}$.
Evidently, among generators of $E_{11}$, commuting with $e_0$, are
$E_{10}$ generators, constructed from elements $e_i,h_i,f_i$ with
indexes starting from 2. Next, among them are all generators of
$E_{11}$ with non-zero power of $f_1$. It remains to consider
generators of $e_{11}$ with non-zero power of $e_1$. They all have
nonzero commutators with $e_0$, which is evident from the rules of
construction of roots - the scalar product of such generators with
$e_0$ are nonzero negative integers (equal to the power of $e_1$),
so real root $\alpha_{0}$ can be added to the given root of
$E_{11}$. So, the stabilizer of $e_0$ within $E_{11}$ is $(E_{10},
(f_1)^n... (n>0))$, where $(f_1)^n... (n>0))$ denotes all roots of
$E_{11}$ with nonzero power of $f_1$. This is a semidirect product
of $E_{10}$ and $(f_1)^n... (n>0)$. So, particularly, each unitary
representation of $E_{10}$ gives rise to induced unitary irrep of
$l_1\ltimes E_{11}$.

\section{Conclusion}

In the body of paper we discuss some features of $E_{11}$
approach, which can help in study of different aspects of theory -
such as orbifolds, supersymmetry relation, induced
representations, etc. We introduce an even subalgebras $Z_2(E_n)$
and find a description of corresponding roots through $EE_{11}$
(for $E_{11}$), $DE_{10}$ (for $E_{10}$), and $D_8^{(1)}$ (for
$E_{9}$). It is proved that for last two cases even roots are
completely given by algebras mentioned, for $E_{11}$ that is a
hypothesis, supported by computer calculations up to level 142.
The possible form of supersymmetry relation (\ref{susy}) is
considered. It requires that compact subgroup $K_n$ has a
representation the symmetric square of which contains $l_1$
representation of $E_{11}$. In view of that we consider the
expansion of space of brane charges, i.e. $l_1$ w.r.t. the
SO(10,10) subgroup of $E_{11}$, and show that first few
representations of $SO(10,10)$ in $l_1$ can be represented in
required form. Corresponding relation (\ref{susy2}) is a standard
11d supersymmetry relation, decomposed w.r.t. the $SO(10) \times
SO(10)$ subgroup. Another result, which may have relation with
supersymmetry (and not only) is the finding of Section 3, that
second fundamental weight of $EE_{11}$ coincides with the weight
of $l_1$ irrep of $E_{11}$. Other existing approaches to
supersymmetrization of $E_{11}$ are discussed.

Finally, precise answer is obtained for a little group of particle
orbit in a semidirect product group $l1\ltimes E_{11}$, assumed to
be an extended symmetry group of $E_{11}$ theory.

In conclusion, we would like to discuss the possible connection of
$E_{11}$  approach to well-known method of coadjoint orbits.
According to the construction of \cite{w0} fields of
$E_{11}/K_{11}$ manifold contains both fields and their duals, as
is precisely shown for lower level, and corresponding equations of
motion are those of generalized self-duality. If one neglects
dependence of fields from (infinite number of?)  space-time
coordinates, that will mean that $E_{11}/K_{11}$ is not a
configuration space but rather phase space, since it includes both
fields and their conjugate momenta. Taking into account the
existence of natural (Kirillov-Kostant) Poisson bracket on the
coadjoint orbits of Lie algebras, one can ask whether
$E_{11}/K_{11}$ is such an orbit. That would mean that $K_{11}$ is
a stabilizer of some element of $E_{11}$ algebra. It is easy to
show that it is not the case. Of course, according to the previous
discussion it is not a necessary feature, one should take into
account a space coordinates, to make a statement precise. We think
that application of coadjoint orbit method to $E_{11}$ worths
further study.

\section{Acknowledgements}

RM is indebted to P.West for discussions during Theoretical
Physics Conference at Lebedev Physics Institute and for e-mail
correspondence. We acknowledge the e-mail correspondence with
H.Nicolai, O.Ganor and I.Schnakenburg. Work is partly supported by
INTAS grant 03-51-6346.

\end{document}